# Earthquake scenarios and seismic input for cultural heritage: applications to the cities of Rome and Florence

**Fabio Romanelli**[a,b], **Franco Vaccari**[a,b,c]

[a] Department of Mathematics and Geosciences, University of Trieste

[b] SAND Group - The Abdus Salam International Centre for Theoretical Physics (ICTP)

[c] XeRiS - eXact lab Srl, Trieste

Keywords: Earthquake scenario; Seismic Hazard, Seismic input.

**INTRODUCTION**

For historical buildings and monuments, i.e. when considering time intervals of about a million year (we do not want to loose cultural heritage), the applicability of standard estimates of seismic hazard is really questionable. A viable alternative is represented by the use of the scenario earthquakes, characterized at least in terms of magnitude, distance and faulting style, also taking into account the complexity of the source rupturing process. Scenario-based seismic hazard maps are purely based on geophysical and seismotectonic features of a region and take into account the occurrence frequency of earthquakes only for their classification into exceptional (catastrophic), rare (disastrous), sporadic (very strong), occasional (strong) and frequent. Therefore they may provide an upper bound for the ground motion levels to be expected for most regions of the world. The neo-deterministic approach naturally supplies realistic time series of ground motion, which represent also reliable estimates of ground displacement readily applicable to seismic isolation techniques, useful to preserve historical monuments and relevant man made structures. This methodology has been successfully applied to many urban areas worldwide for the purpose of seismic microzoning, to strategic buildings, lifelines and cultural heritage sites. We will discuss its application to the cities of Florence and, more extensively, Rome.

## THE NEODETERMINISTIC SEISMIC HAZARD ASSESSMENT (NDSHA) AT THE LOCAL SCALE

By neo-deterministic we mean an innovative and widely applied deterministic procedure (e.g. Panza et al. 2001a; 2012), that supplies realistic time histories from which it is possible to retrieve, in correspondence of earthquake scenarios: a) peak values for ground displacement, velocity and design acceleration at bedrock level for the regional scale, and b) other relevant damage indicators for the local scale.

A proper evaluation of the seismic hazard, and of the seismic ground motion due to an earthquake, can be accomplished by following a deterministic or scenario-based approach, coupled with engineering judgment. This approach allows us to incorporate all available information collected in a geological, seismotectonic and geotechnical database of the site of interest as well as advanced physical modelling techniques to provide a reliable and robust basis for the development of a deterministic design basis for cultural heritage and civil infrastructures in general.
Where the numerical modelling is successfully compared with records, the synthetic seismograms permit the microzoning, based upon a set of possible scenario earthquakes. Where no recordings are available the synthetic signals can be used to estimate the ground motion without having to wait for a strong earthquake to occur (pre-disaster microzonation).
One of the most difficult tasks in earthquake scenario modelling is the treatment of uncertainties, since each of the key parameters has an uncertainty and natural variability, which often are not quantified explicitly. A possible way to handle this problem is to vary the modelling parameters systematically. Actually, a severe underestimation of the hazard could come by fixing a priori some source characteristics and thus the parametric study should take into account the effects of the various focal mechanism parameters (i.e. strike, dip, rake, depth etc.).
The analysis of the parametric studies allow to generate advanced groundshaking scenarios for the proper evaluation of the site-specific seismic hazard, with a complementary check based on both probabilistic and empirical procedures.
Once the gross features of the seismic hazard are defined, and the parametric analyses have been performed, a more detailed modelling of the ground motion can be carried out for sites of specific interest. Such a detailed analysis should take into account the source characteristics, the path and the local geological and geotechnical conditions.
In Figure 1 the flowchart of the approach is shown.

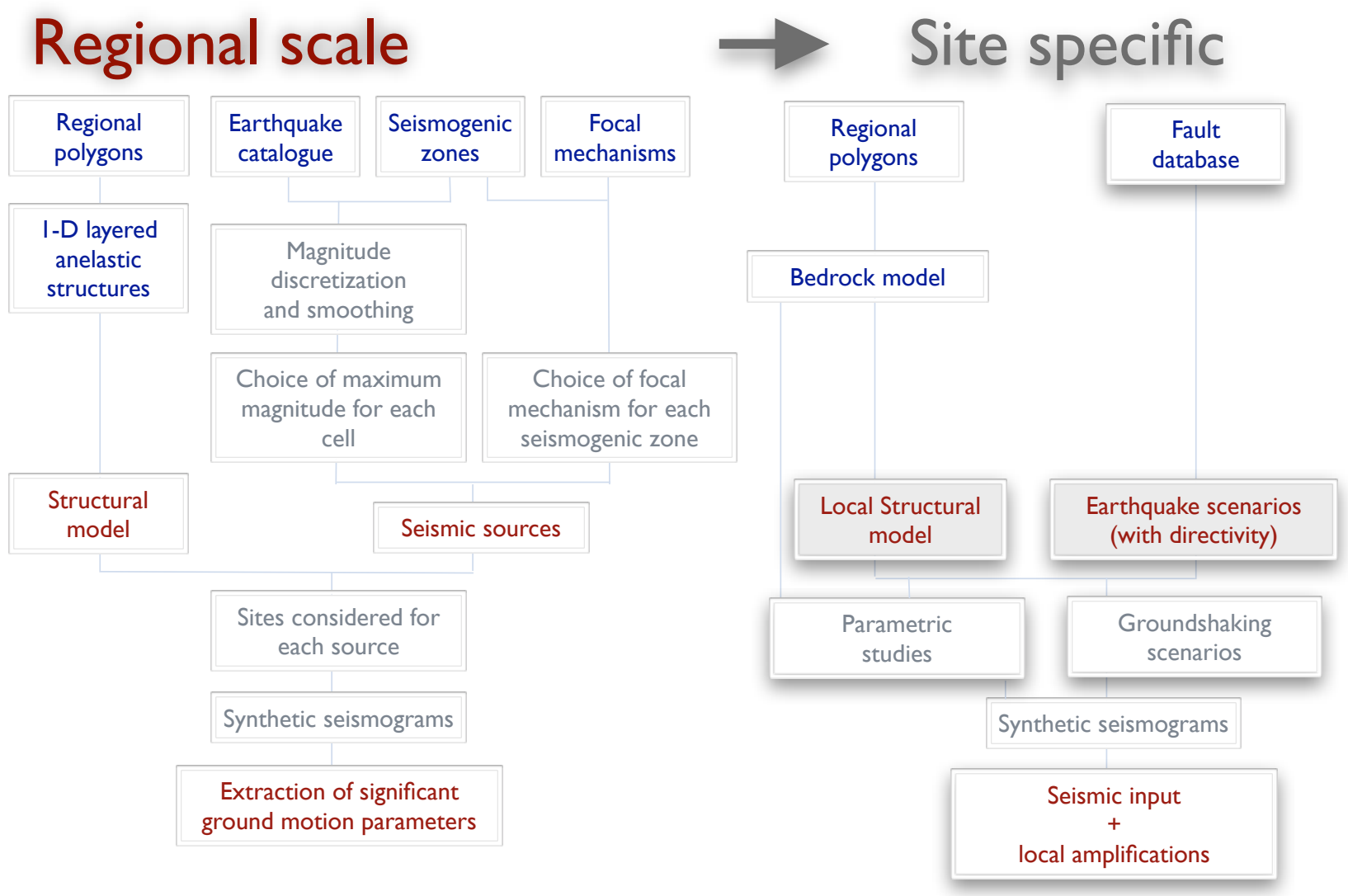

*Figure 1. Flowchart of the site specific NDSHA approach*

This methodology has been successfully applied to many areas worldwide (e.g. Panza et al. 2001a,b; 2012), to strategic buildings, lifelines and cultural heritage sites, and, for the purpose of seismic microzoning, to several urban areas, in the framework of the UNESCO/IUGS/IGCP project "Realistic Modelling of Seismic Input for Megacities and Large Urban Areas" (e.g. Panza et al. 200b).

The approach has been successfully applied also to the city of Florence, where Negro (2013) performed a site-specific vulnerability study for the "Prigioni" statues. She estimated the seismic input and evaluated risk mitigation techniques for cultural heritage. In this work we simply show (Figure 2) an example of parametric study for the scenario earthquake on the Prato-Fiesole fault, performed with the XeRiS web application (Vaccari, 2016).

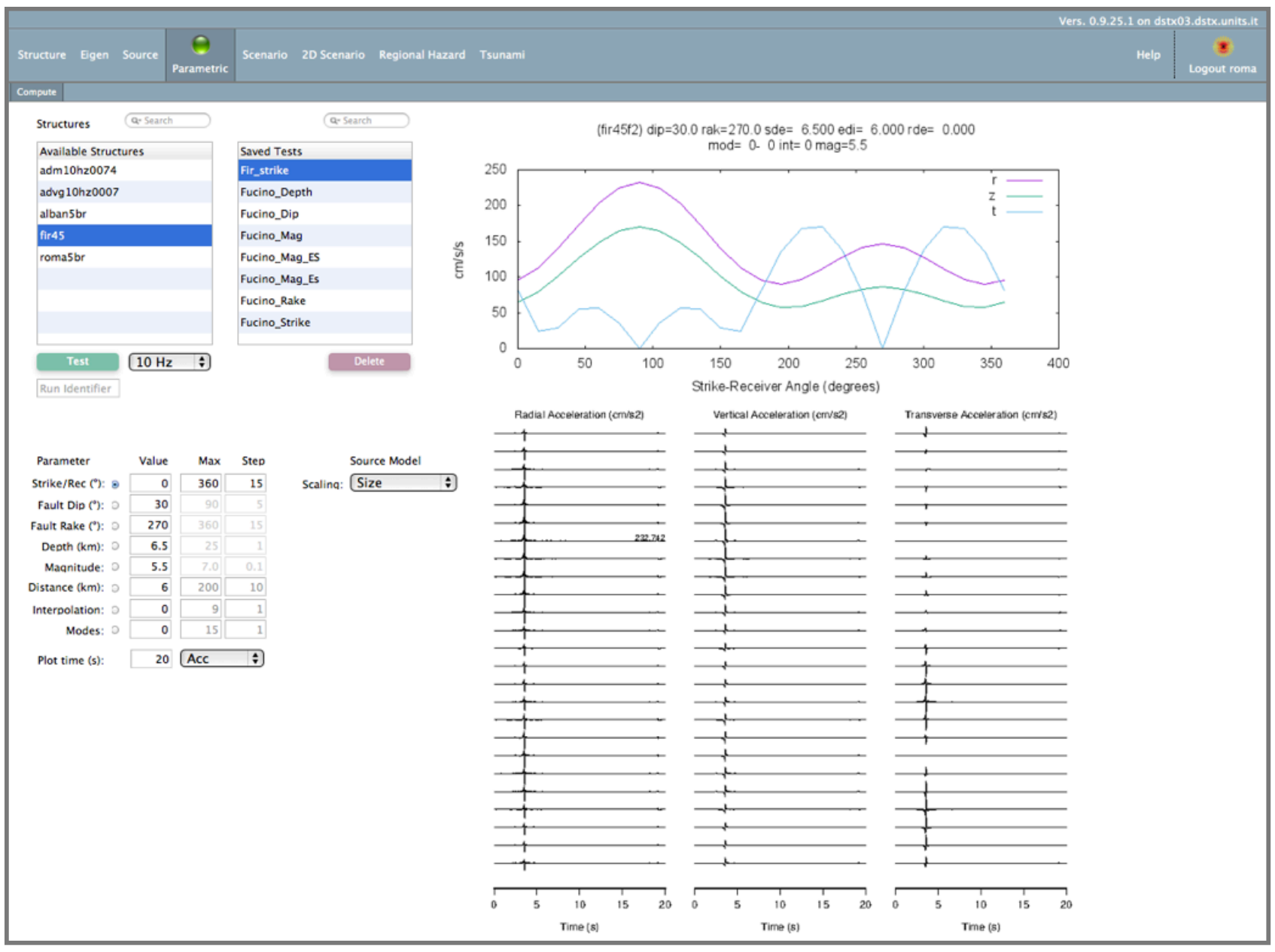

*Figure 2. Example of parametric study (strike-receiver angle) for the scenario earthquake on the Prato-Fiesole fault*

**Microzonation of Rome**

This section aims to represent an update of the various contributions devoted to this topic (e.g. see Fäh et al. 1993 and references therein). The well-documented distribution of damage in Rome caused by the January 13, 1915 Fucino earthquake - (Intensity XI on the Mercalli-Cancani-Sieberg, MCS, scale) was successfully compared with the reported macroseismic intensities by Fäh et al. (1993; 1995). They extended (Figure 3) the zoning to the entire city of Rome, providing a basis for the prediction of the expected damage from future strong events.

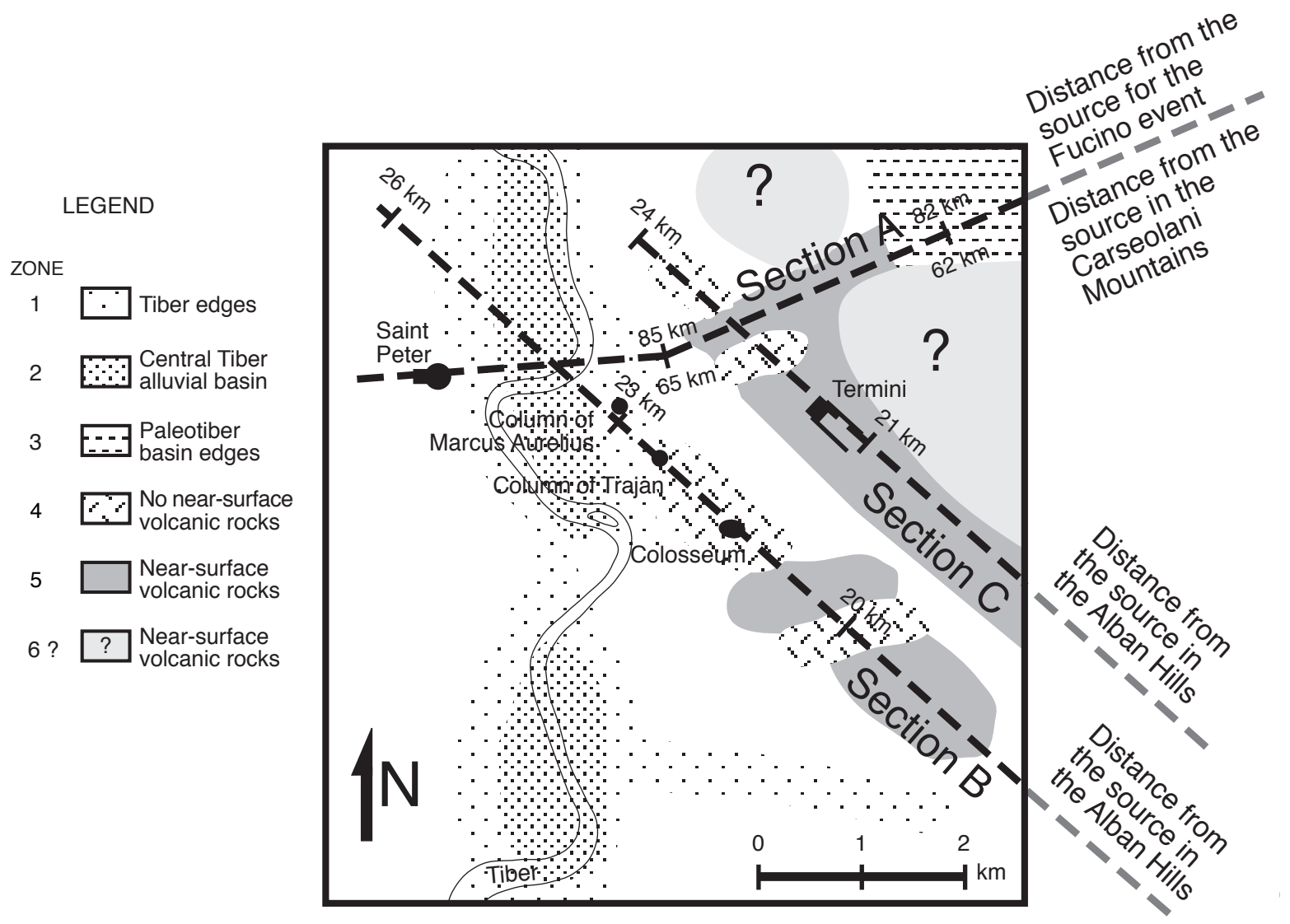

*Figure 3. Microzonation map of Rome (from Fäh et al., 1995)*

We reconsidered two of the seismogenic zones around Rome: the Fucino area and the Alban Hills, performing parametric studies starting from the related information contained in the DISS (2010) and SHARE (2014) databases. As an example the groundshaking scenario, computed at bedrock, is shown in Figure 4. To deal both with realistic source and local structural models, we used the hybrid method that combines the modal summation and the finite difference technique (e.g. Fäh et al. 1993), optimizing the use of the advantages of both methods. Source, path and site effects are all taken into account, and it is therefore possible a detailed study of the wavefield that propagates even at large distances from the epicentre. An example of the acceleration time series obtained along the profile A (see Figure 3) for the Fucino scenario is shown in Figure 5.

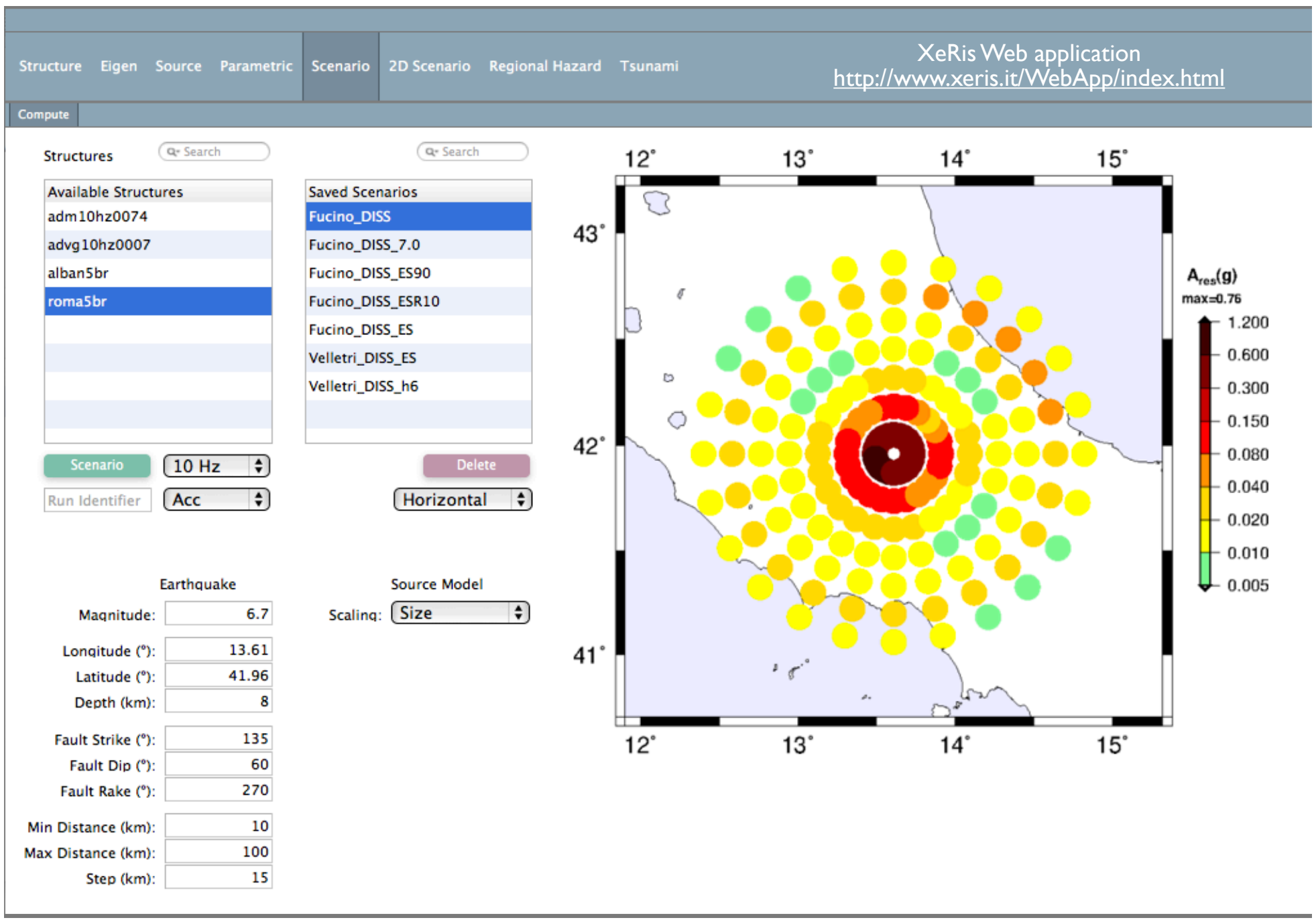

Figure 4. Groundshaking scenario for the Fucino (1915) event

Assuming a realistic kinematic description of the rupture process, the stochastic structure of the accelerograms can be reproduced. The extended seismic source model allows us to generate a spectrum (amplitude and phase) of the source time function that takes into accounts both the rupture process and the directivity effects (Gusev 2011). We sum up the source time functions generated by the distributed (point) subsources in order to obtain the equivalent single source, representative of the entire space and time structure of the extended source, and the related Green's Function. In such a way it is possible to perform expeditious parametric studies, useful for engineering analysis.

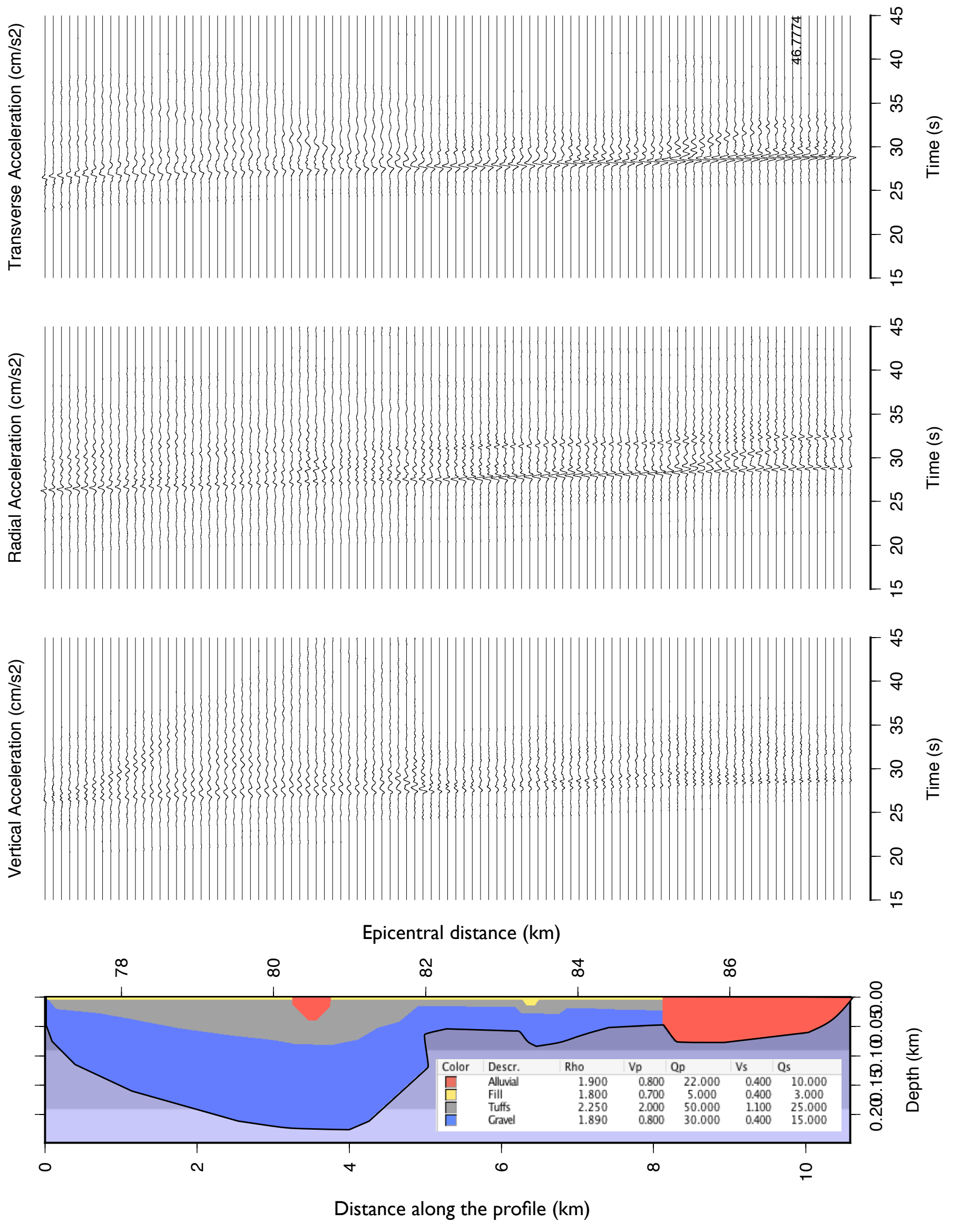

Figure 5. acceleration time series obtained along the profile A (see Figure 3) for the Fucino scenario.

We applied this technique to estimate the uncertainty associated with the source effect, considering two specific sites (at 87 km and 19 km respectively) along profiles A and B, for the Fucino and Colli Albani scenarios. The results for the spectral accelerations are shown in Figure 6, compared with NTC08 (2008) spectra (with a return period of 475 and 2500 years) provided for the node around Rome (ID 28375) with highest PGA value.

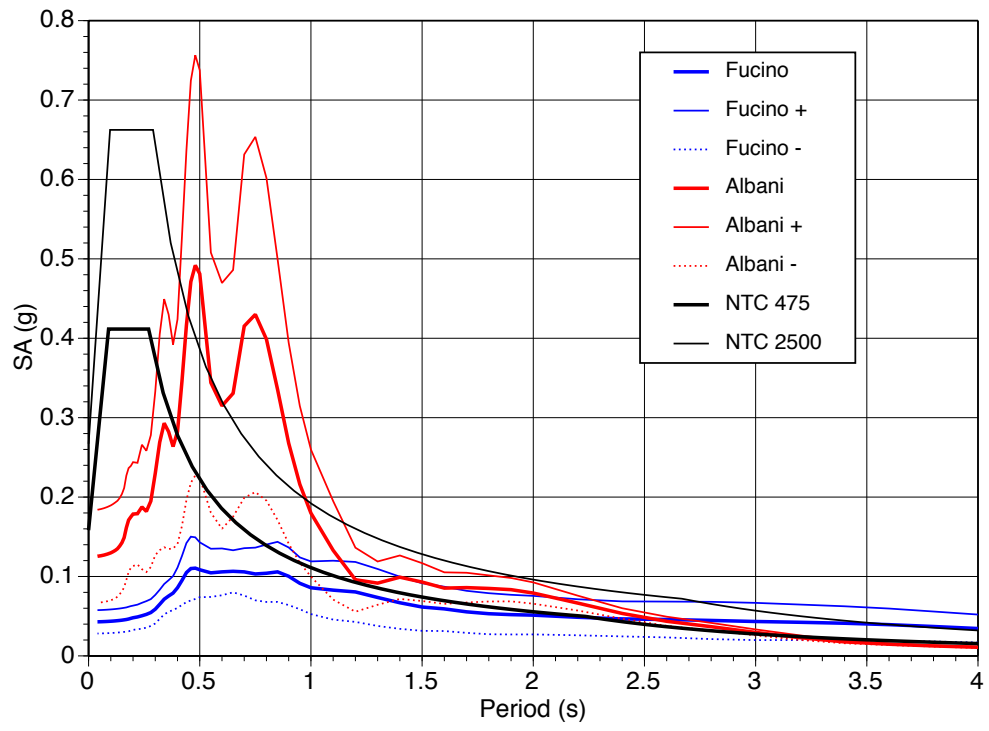

Figure 6. Blue lines: response spectra for the site at 87km epicentral distance along profile A from the Fucino scenario, plus and minus one standard deviation (+,-); Red lines: same for the site at 19km epicentral distance along profile B from the Colli Albani scenario; Black lines: NTC08 spectra (for 475 and 2500 return period).

The results obtained from the parametric study for the kinematic parameters of the seismic sources can be used to estimate the possible related uncertainty associated with the six spectral shapes defined for the microzoning of Rome (see Figure 3; Fäh et al. 1993; Panza et al., 2000). The results are shown In Figure 7.

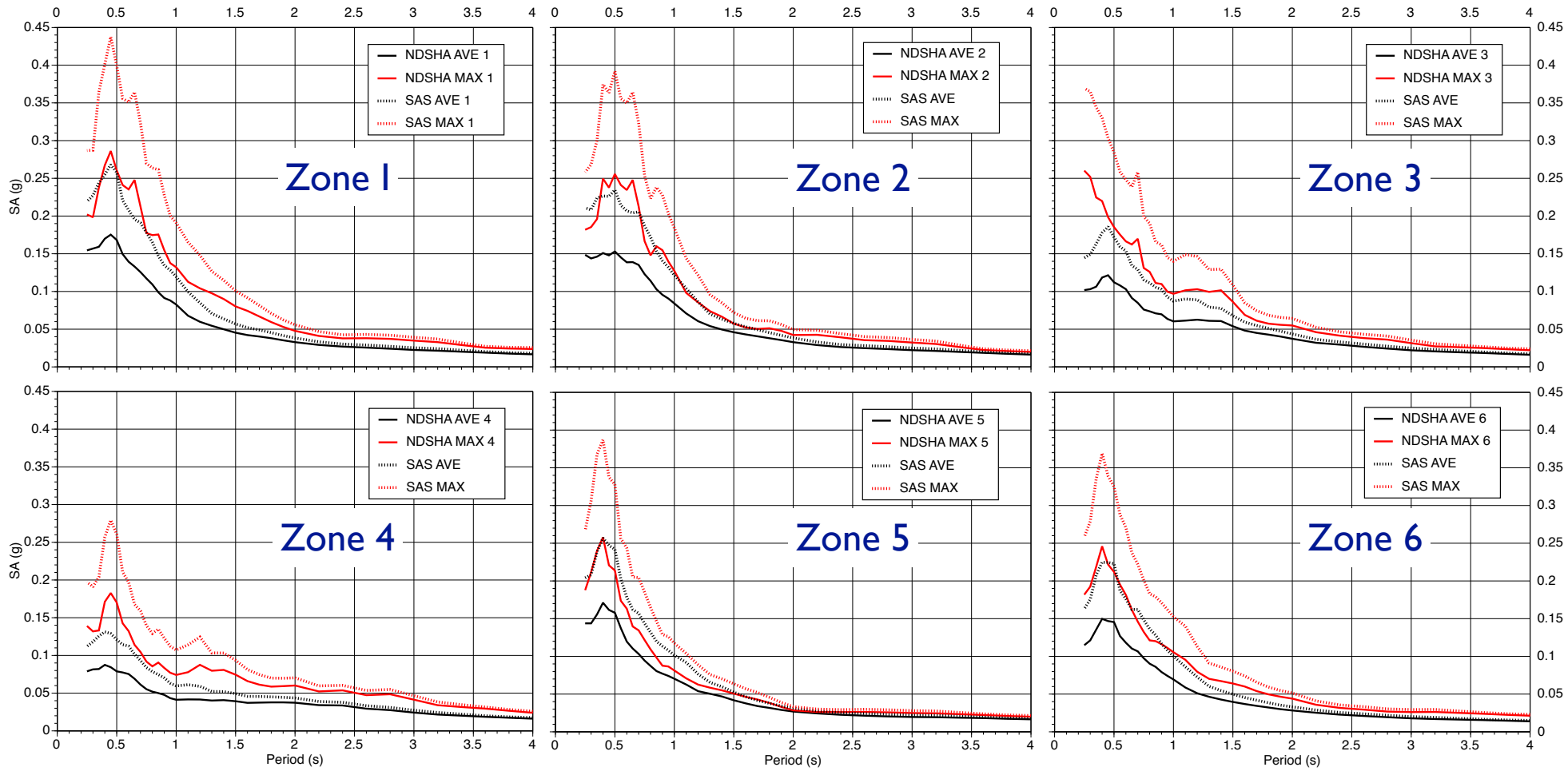

Figure 7. For the six zones identified in Figure 3, the maximum (NDSHA MAX)and average (NDSHA AVE) absolute spectral accelerations are taken from Peresan et al. (2004). The estimate of the 84th percentile related to the source and site effects (SAS) is represented by the dashed lines

The results, that are in agreement with the spectral analysis carried out recently by complementary approaches (e.g. Sabetta, 2014), confirm: a) the combined importance of site and source effects on the characterisation of seismic input, and b) that NDSHA naturally supplies realistic time series of ground motion, which represent also reliable estimates of ground displacement readily applicable to seismic isolation techniques, useful to preserve historical monuments and relevant man made structures.